\documentclass[prl, twocolumn, groupedaddress, showpacs, showkeys]{revtex4}
\usepackage[]{natbib}

\begin{document}

\title{Quantum Statistics and Entanglement Problems}

\date{\today}

\author{L. E. H. Trainor}
\affiliation{Department of Physics,
            University of Toronto,
            Toronto, Canada M5S 1A8}

\email[Correspondence:]{ltrainor@physics.utoronto.ca}
\thanks{Dept. of Physics, University of Toronto, 180 St. George St., Toronto, Canada M5S 1A8}

\author{Charles J. Lumsden}
\affiliation{Institute of Medical Science,
            University of Toronto,
            Toronto, Canada M5S 1A8}

\begin{abstract}
Interpretations of quantum measurement theory have been plagued by two questions,
one concerning the role of observer consciousness and the other the entanglement
phenomenon arising from the superposition of quantum states. We emphasize here the remarkable 
role of quantum statistics in describing the entanglement problem correctly and discuss
the relationship to issues arising from current discussions of intelligent observers
in entangled, decohering quantum worlds. 
\end{abstract}

\pacs{03.65.Ta, 03.65.Ud, 03.65.Yz 05.30.Fk, 87.19.Bb, 01.70+w}

\keywords{entanglement, spin, quantum statistics, decoherence, einselection, Many Worlds Interpretation}

\maketitle

\section{Introduction}
In the Copenhagen interpretation of quantum mechanics and the theory of 
measurement associated with it, one is required a priori to assume the existence of a 
classical measuring apparatus. To many physicists this has appeared to be a 
circular argument because the classical apparatus itself is composed of quantum particles, 
so why is not a quantum mechanical description possible ``all the way down"? Recently 
great attention has been devoted to this problem \cite{ZehItBit, ZehQM, TegShap1994, JoosDecoh, ZurekDecoh, TegWheeler} through the 
supposition of environmental influences, such as the assumption of a global, all 
encompassing Schr\"odinger wave function. In this theory, classical mechanics is 
recovered from a quantum description by means of decoherence and einselection \cite{ZurekDecoh}. It is 
generally recognized that a basic problem to be resolved in this process is the one arising 
from the superposition principle of quantum mechanics and its resulting entanglement  
complication.

The purpose of this letter is to stress the complications for 
the interpretations in quantum mechanics of many-particle wave functions, first studied 
seriously by Pauli \cite{Pauli57}, and to re-emphasize the role of quantum statistics, in distinction to dynamics, 
in the quantum entanglement problem.

 \section{Pauli Entanglement}
Both in the light of Pauli's remarks and with an eye toward recognizing  the 
importance of the Bell Theorem \cite{Bell1964, Aspect1982} in showing that EPR (Einstein, Podolsky and 
Rosen) were unjustified in claiming \cite{EPR1935} that any complete description of reality must
include a locality assumption, we use a simple 
case of a two-particle wave function to illustrate the two points made in the previous 
paragraph.

	First let us consider a general, two-particle  wave function ({\it non}-identical particles) ${\psi}(x_1,x_2)$, isolated
with coordinates $x_1$ and $x_2$ 
 with time evolution implicit.  In the interest of exploring EPR 
separability, we suppose that during some portion of the two-particle history, this wave 
function separates into a product of two single particle wave functions, one dependent on  $x_1$, the other on  $x_2$:
 \begin{equation}
		{\psi}(x_1,x_2)  =   \phi(x_1)\chi(x_2)					\label{Eq:psi12}
\end{equation}

\noindent
and we enquire into the interpretation of $\phi(x_1)$. We do this by multiplying both sides of 
Eq.~(\ref{Eq:psi12}) by  $\chi^*(x_2)$ and integrating over the coordinates of  $x_2$   (assuming normalized 
single particle wave functions) to obtain
\begin{equation}
		  \phi(x_1)    =    \int {\psi}(x_1,x_2) \chi^*(x_2) dx_2  \thinspace					\label{Eq:phi_x1} 
\end{equation}
	
\noindent
This familiar procedure makes it clear that even in product form, the wave function for particle 1 
is entangled with that of  particle 2, and that the criterion for locality  required by  EPR is not 
realized since a measurement on particle 2 at any time changes  $\chi^*(x_2)$ and thus the 
integral in Eq.~(\ref{Eq:phi_x1}), so that  $\phi(x_1)$ is not a true description of particle 1 
independently of what happens to particle 2. 

The entanglement situation is even more interesting, however, when one considers two {\it identical} fermions, since then one has the 
Pauli requirement that  ${\psi}(x_1,x_2)$ in Eq.~(\ref{Eq:psi12}) be replaced by
\begin{equation}
		   {\Psi}_A(x_1,x_2)   =   {1 \over \sqrt{2}}[ \phi(x_1)\chi(x_2)  -  \phi(x_2)\chi(x_1) ]	  	\label{Eq:Pauli_1} 
\end{equation}

\noindent
We see that entanglement is profoundly enhanced by the Pauli principle since the simplifying 
procedure leading to Eq.~(\ref{Eq:phi_x1} ) no longer holds. We further note that any 
generalization to Eq.~(\ref{Eq:psi12}) such as
\begin{equation}
        {\psi}(x_1,x_2)  =    \sum_{i.j} C_{ij} \phi_i(x_1) \phi_j(x_2)					\label{Eq:Pauli_2}
\end{equation}

\noindent
where the $C_{ij}$ are $C$-numbers, will have similar properties, particularly when the 
requirement of antisymmetry is imposed.

In the case of two identical bosons the same argument leads to similar results except a plus 
sign replaces the minus sign on the right hand side of Eq.~(\ref{Eq:Pauli_1}), with
\begin{equation}
		   {\Psi}_S(x_1,x_2)   =   {1 \over \sqrt{2}}[ \phi(x_1)\chi(x_2)  +  \phi(x_2)\chi(x_1) ] \thinspace .   \label{Eq:Bosons_1}
\end{equation}

\noindent
Supersymmetry has no consequence for these arguments since we live in a 
world where supersymmetry has already been broken.

\section{Global Entanglement}

Insufficient attention has been paid to entanglement due to 
statistics, particularly in view of the current interest expressed in a global  Schr\"odinger 
wave function and pointer states \cite{JoosDecoh, ZurekDecoh}.  Decoherence due to the ``environment" is used to 
suggest that classical systems can be deduced from overarching quantum 
systems  \cite{TegShap1994, TegWheeler}, thus avoiding the need in the Copenhagen interpretation for a separate 
classical world (the apparatus and observer) in the measurement process, joined somehow by the
punctate collapse of the wave packet.
 
It is worth emphasis that the symmetrization and antisymmetrization of wave 
functions for identical bosons, respectively fermions, is not inherent in the dynamics of 
the Schr\"odinger wave function theory, but must be imposed outside the dynamical 
theory itself. There is no dynamical procedure in this picture, for example, that tells two electrons that 
they must antisymmetrize at some point in their mutual histories.  The Pauli principle 
itself has the effect of entangling the wave functions of all electrons in the universe.  This   
entanglement can often be ignored because the electrons ``outside" the system are separated well beyond their 
dynamical range of interaction with electrons ``inside" the system. Second quantization, which is
designed to deal with many particle systems, does not alleviate this situation, since the
use of commutators, repseectively anticommutators, leads to results in consonance with the
first quantization picture in these respects, namely that once particles come within the dynamical
range of their interactions, the symmetrization/antisymmetrization behavior must be recognized and
taken into account.  

 This aspect of quantum 
mechanics is well illustrated in the experiments described by Mott and Massey \cite{Mott1965} 
in which electrons in a
monochromatic beam were scattered off  hydrogen atoms in their ground state, and the angular distributions observed. The results 
were analyzed and explained by  Trainor and Wu  (\cite{TW1953}; TW hereafter) and by Corinaldesi and Trainor (\cite{CT1952}; CT
hereafter).   TW showed that to obtain the experimental results antisymmetrization of the incident and target electrons had to be taken
into account. This could be done by antisymmetrization of the electron pairs in their initial states, with the 
antisymmetrization then maintained throughout the analysis. CT subsequently established that in the Born-Oppenhemier approximation all of
the required interaction matrices could be calculated exactly. The Schr\"odinger dynamics only gives the correct results when
antisymmetrization of the beam electrons and the target electrons is imposed on the formation of the initial states.
An electron in the incoming beam could have arrived from Japan and collided with a hydrogen electron in American for all the
system cares.

\section{Pauli, Bell, and MWI: No Safe Harbors}

Once it became clear from Bell's theorem \cite{Bell1964} and the experimental verification of quantum mechanics \cite{Aspect1982}
that EPR's assumption was incorrect --- reality did {\it not} demand local behavior for a complete theory --- 
there were influential attempts to minimize the impact of these results. Ballentine, for example, proposed \cite{Bal1970} 
that the statistical interpretation of quantum mechanics brought it into closer accord with an EPR-type
reality than could be realized from a single particle point of view. This is not the case, however.
Kunstatter and Trainor (\cite{KT1984A}; KT hereafter) showed that in the statistical interpretation, the results of Bell \cite{Bell1964}
and Aspect \cite{Aspect1982} also required EPR-like locality assumptions to be abandoned.

Another important attempt to contain the non-locality issue was made by Page \cite{Page1982} using the Many Worlds Interpretation
(\cite{MWI1957}; MWI hereafter). KT  \cite{KT1984B} argued, however, that the ideas used to describe observers in quantum measurement
theory are largely intuitions based on what cognitive scientists and philosophers of mind now term ``folk psychology"
\cite{Churchland1986} and thus highly problematic in their own right: ``do intelligent observers exist in quantum mechanics?"
\cite{KT1984B}. The nature and role of conscious observers in quantum mechanics, particularly as raised by Wigner \cite{Wigner1962}, has
continued to plague and challenge interpretation theories. The challenges are even greater for theories based on the proposition that
there is a global Schr\"odinger equation for the Universe as a whole, and that it is ``quantum mechanics all the way down", particularly
since there is no accepted quantum (or any other) theory of consciousness. This point is emphasized, for example, in the recent exchange
between Tegmark
\cite{TegBrain} and Hagan et al. \cite{HaganQMBrain} on quantum decoherence issues relating to the role of neuronal microtubules as basic
quantum computational elements in a theory of self awareness, subjectivity, and memory in the human brain.

The recent paper of Zeh \cite{ZehQMCs} proposing that, when worlds divide in the Everett's theory \cite{MWI1957}, minds also divide
is a bold and speculative attempt to deal with such questions as decoherence and MWI, but leaves untouched those raised by KT on whether
intelligent behavior as such is even describable by the quantum theories we now have. Indeed, to conjecture in the affirmative is to
assert a locality hypothesis of a different sort within quantum mechanics, namely that mind, self-awareness, and perception
as phenomena may be inferred from the theory, i.e. traced to a finite (and thus circumscribed or epistemically ``local") set of 
assumptions (the only kind of axiom set we currently know how to reason with). Since local realism --- EPR-like locality
requirements on either the single or quantum statistical levels --- has fared so poorly to date, it seems reasonble to conjecture that
no safe local harbors exist at all within quantum theory, and that entanglement and decoherence point to problems about observers
and the Universe whose
solution will entail further significant changes in our understanding of quantum mechanics.

\end{document}